\documentclass[
reprint, 
 amsmath,amssymb,
 aps, prd,twocolumn,
]{revtex4}
\usepackage{subfigure}
\usepackage{caption2}
\usepackage{graphicx}
\usepackage{dcolumn}
\usepackage{amsmath}
\usepackage{amssymb}
\usepackage{bm}
\usepackage{hyperref}
\usepackage{xcolor}
\usepackage[mathlines]{lineno}

\usepackage{slashed}
\begin{document}


\title{\textbf{Study of the $\Omega_{ccc}\Omega_{ccc}$ and $\Omega_{bbb}\Omega_{bbb}$ dibaryons in QCD Sum Rules} 
}%

\author{Xu-Liang Chen$^1$}
\author{Jin-Peng Zhang$^1$}
\author{Zi-Xi Ou-Yang$^2$}
\author{Wei Chen$^{1,\, 3}$}
\email{chenwei29@mail.sysu.edu.cn}
\author{Jia-Jun Wu$^{2,\, 3}$}
\email{wujiajun@ucas.ac.cn}

\affiliation{$^1$School of Physics, Sun Yat-sen University, Guangzhou 510275, China \\ 
$^2$School of Physical Sciences, University of Chinese Academy of Sciences, Yuquan Road 19A, Beijing 100049, China \\
$^3$Southern Center for Nuclear-Science Theory (SCNT), Institute of Modern Physics, 
Chinese Academy of Sciences, Huizhou 516000, Guangdong Province, China}

\begin{abstract}
The recent observation of a family of fully-charm tetraquark states by the LHCb, ATLAS and CMS Collaborations suggests the possible existence of fully-heavy dibaryons. In this work, we investigate the $\Omega_{ccc}\Omega_{ccc}$ and $\Omega_{bbb}\Omega_{bbb}$ dibaryons in both the $^1S_0$ and $^5S_2$ channels using the method of QCD sum rules. We employ the iterative dispersion relation (IDR) method to efficiently compute the massive five-loop banana diagrams that appear in these systems, and properly address the tricky small-circle divergence problem in the nonperturbative terms. Our analyses reveal that for both charm and bottom systems, the scalar dibaryon lies lower than its tensor counterpart. In $\overline{\text{MS}}$ scheme, the mass of the scalar $\Omega_{ccc}\Omega_{ccc}$ dibaryon is found to be slightly above the $2\Omega_{ccc}$ mass  threshold, while the $\Omega_{bbb}\Omega_{bbb}$ systems may form bound states. However, they are predicted to be much heavier in the on-shell scheme.
\end{abstract}

\maketitle

\section{Introduction}
The baryon-baryon interaction is a notable topic in hadron physics. It determines the fundamental properties of atomic nuclei and serves as a key anchor for exploring the nature of strong interactions. A dibaryon is the simplest platform for studying the baryon-baryon interaction. Experimentally, the only confirmed dibaryon to date is the deuteron. Discovered in 1932, it has a binding energy of approximately $2.2\,\mathrm{MeV}$~\cite{PhysRev.39.164,PhysRev.40.1}. Another possible candidate for dibaryon is the recently discovered $d^*(2380)$ by the WASA-at-COSY Collaboration~\cite{WASA-at-COSY:2011bjg,WASA-at-COSY:2013fzt}, which could be a bound state of $\Delta \Delta$~\cite{Dyson:1964xwa}. More details can be found in Refs.~\cite{Clement:2016vnl,Gal:2015rev,Gal:2016eqi}.

Since 2020, a family of $X(6600), X(6900), X(7200)$ states has been observed in succession by the LHCb, ATLAS, and CMS Collaborations through their decays into the $J/\psi J/\psi$ and $J/\psi\psi(2S)$ final states~\cite{LHCb:2020bwg,CMS:2023owd,ATLAS:2023bft}. As candidates of fully-charm tetraquark states, their observations inspired the existence of fully-heavy dibaryons, which represent exceptionally distinctive and intriguing states within the hadron spectrum. Due to the large mass of their constituent quarks, relativistic effects can be safely neglected, which significantly simplifies their treatment within potential quark models. Moreover, the absence of light quarks in these systems strictly forbids the exchange of light mesons (e.g., $\pi$, $\sigma$, $\rho$, and $\omega$). Consequently, the interaction between the two heavy hadrons is mediated exclusively by heavy quarkonia, offering a unique window into the dynamics of hadron-hadron interactions in a pristine, light-quark-free environment.

To date, various theoretical analyses have been conducted on fully-heavy dibaryons. The first Lattice QCD study of the $\Omega_{ccc}\Omega_{ccc}$ system in the $^1S_0$ channel was performed by the HAL QCD Collaboration, which suggested the formation of a loosely bound state~\cite{Lyu:2021qsh}. Subsequently, Mathur et al. carried out an extensive investigation of heavy-flavor dibaryons, providing a comprehensive set of calculations that span multiple systems and methodologies~\cite{Mathur:2022ovu,Dhindsa:2025gae,Junnarkar:2019equ,Junnarkar:2022yak,Junnarkar:2024kwd}. For fully heavy dibaryons, their results consistently reveal a predominantly negative energy shift $\Delta E = E_D - 2E_B$, indicating a systematic binding trend across the channels studied.

The quark model has also been extensively employed in the investigation of fully-heavy dibaryons~\cite{Martin-Higueras:2024qaw,Huang:2020bmb,Weng:2022ohh,Lu:2022myk,Alcaraz-Pelegrina:2022fsi,Richard:2020zxb,Gordillo:2023tnz}. However, conclusions drawn from these studies remain a subject of debate. While some model calculations support the existence of bound states~\cite{Martin-Higueras:2024qaw,Huang:2020bmb}, while others find no evidence for binding~\cite{Weng:2022ohh,Lu:2022myk,Alcaraz-Pelegrina:2022fsi}. Complementary approaches have yielded further insights. For instance, the One-Boson Exchange (OBE) model was employed in Ref.~\cite{Liu:2021pdu} to study the $\Omega\Omega$, $\Omega_{ccc}\Omega_{ccc}$, and $\Omega_{bbb}\Omega_{bbb}$ systems, concluding that all three dibaryons form bound states. Notably, that study also highlighted that the Coulomb interaction may be sufficient to dissociate the $\Omega_{ccc}\Omega_{ccc}$ pair, while the $\Omega\Omega$ and $\Omega_{bbb}\Omega_{bbb}$ systems remain bound. Within the framework of QCD sum rules, Ref.~\cite{Wang:2022jvk} investigated the compact diquark-diquark-diquark fully-heavy hexaquark states and found that all these hexaquarks lied below the corresponding two baryons mass thresholds. A comprehensive summary of the predictions from these various theoretical approaches is presented in Table~\ref{table:comparison with others}.
In this work, we shall consider the molecular currents to study $\Omega_{ccc}\Omega_{ccc}$ and $\Omega_{bbb}\Omega_{bbb}$ dibaryons in scalar $^1S_0$ and tensor $^5S_2$ channels in the QCD sum rules method~\cite{Shifman:1978bx,Shifman:1978by,Reinders:1984sr}. 

This paper is organized as follows. In Sec.~\ref{sec2}, we present the QCD sum rules for dibaryon states and describe the mathematical techniques developed to handle the complicated massive five-loop Feynman diagrams, with particular attention to the challenging small-circle divergence problem. In Sec.~\ref{sec3}, we perform the numerical analysis and extract the hadron masses of the $\Omega_{ccc}\Omega_{ccc}$ and $\Omega_{bbb}\Omega_{bbb}$ dibaryons. Finally, Sec.~\ref{sec4} provides a brief summary and conclusions.

\section{Formalism} \label{sec2}
We consider the following interpolating current to describe the fully-heavy dibaryon states in a hadron molecular configuration~\cite{Chen:2019vdh}
\begin{equation}
    \begin{aligned}
    J_{\mu \nu}(x)= & \epsilon^{i_1 j_1 k_1} \epsilon^{i_2 j_2 k_2} [Q^T_{i_1}(x) \mathcal{C} \gamma_\mu Q_{j_1}(x)]Q^T_{k_1}(x)   \\
 & \cdot \mathcal{C} \gamma_5 \cdot Q_{k_2}(x) [Q^T_{i_2}(x) \mathcal{C} \gamma_\nu Q_{j_2}(x)]\, ,
    \end{aligned}
\end{equation}
where $Q=b/c$ denotes a heavy quark field, $(i,j,k)_{1,2}$ are color indices, and the charge conjugation matrix $\mathcal{C}=\mathrm{i}\gamma_2 \gamma_0$. The two-point correlation function in $d$-dimension induced by this interpolating current is defined as
\begin{equation}
    \begin{aligned}
		\Pi_{\mu \nu , \rho \sigma} (q^2) = &\mathrm{i} \int \mathrm{d}^d x \ \mathrm{e}^{\mathrm{i}  q \cdot x} \langle 0 |
		T [J_{\mu \nu} (x) J_{\rho \sigma}^{\dagger} (0)] | 0 \rangle\, . \label{Eq:correlationfunction}
\end{aligned}
\end{equation}
Noting that $J_{\mu \nu}(x)$ is a symmetric operator, it can couple to both the scalar and tensor dibaryons $X_0$ and $X_T$ via the following relations respectively
\begin{align}
\langle0|J_{\mu\nu}|X_0\rangle&=f_{0}g_{\mu\nu}+f_qq_\mu q_\nu\,, \label{scalarcoupling}\\
\langle0|J_{\mu\nu}|X_T\rangle&=f_{T} \epsilon_{\mu\nu}\, ,
\label{tensorcoupling}
\end{align}
in which $f_0, f_q$, and $f_T$ are the coupling constants, and $\epsilon_{\mu\nu}$ is the polarization tensor to the spin-2 state. Accordingly, the correlation function $\Pi_{\mu \nu , \rho \sigma} (q^2)$ contains both invariant functions with $J^P=0^+$ and $2^+$, which can be separated by using the following projectors~\cite{Chen:2011qu,Chen:2017rhl,Chen:2017dpy}
\begin{align}
		& P_{0T} = \frac{1}{16} g_{\mu \nu} g_{\rho \sigma} \, , \qquad  \text{for} \, J^P=0^+ ,\, \text{T}   \nonumber \\
        & P_{0S} = T_{\mu \nu} T_{\rho \sigma} \, , \qquad  \text{for} \, J^P=0^+ ,\, \text{S} \label{eq:projector} \\
        & P_{0TS} = \frac{1}{4} (T_{\mu \nu} g_{\rho \sigma}+T_{\rho \sigma} g_{\mu \nu}) \, , \quad  \text{for} \, J^P=0^+ ,\, \text{TS} \nonumber \\
        & P_{2S} = \frac{1}{2} ( \eta_{\mu \rho} \eta_{\nu \sigma} + \eta_{\mu \sigma} \eta_{\nu \rho} - \frac{2}{3} \eta_{\mu \nu} \eta_{\rho \sigma} ) \, , \quad  \text{for} \, J^P=2^+ ,\, \text{S} \nonumber 
\end{align}
where
\begin{equation}
    \begin{aligned}
		\eta_{\mu \nu}=\frac{q_{\mu}q_{\nu}}{q^2} - g_{\mu \nu} \, ,  T_{\mu \nu}=\frac{q_{\mu} q_{\nu}}{q^2} - \frac{1}{4}g_{\mu \nu} \, .
\end{aligned}
\end{equation}
The projectors $P_{0T},\, P_{0S}, \, P_{0TS}$ in Eq.~\eqref{eq:projector} can pick out the scalar invariant functions with different coupling constants.

Since $J_{\mu\nu}(x)$ contains six identical heavy quark fields, the Wick contraction in Eq.~\eqref{Eq:correlationfunction} will generate 720 terms and make the calculation extremely complicated. We adopt the following procedures to simplify the calculations. Firstly, two terms have the same structure if they coincide after a uniform relabeling: replacing all propagators with a generic symbol $S$, setting color factors to unity, and substituting all Dirac matrices
$ \gamma_{\mu}$ with a generic label $\gamma_a$, ignoring their original indices. Among such terms, permutations of the color and Lorentz indices are tested to find matching expressions for preliminary simplification. Secondly, we say terms have an equivalent structure if, by applying all possible permutations involving translation (cyclic shift) and transpose of Dirac algebra within trace operations, they can be brought into an identical structural form. Through this process, we obtain 36 distinct terms in the final correlation function, significantly reducing the computational time.

At the phenomenological side, the correlation function can be represented via the dispersion relation (DR) as
	\begin{eqnarray}
		 \Pi^{\text{PH}} (q^2) =  \int_{s_N}^{\infty} \mathrm{d} s \frac{ (q^2)^n\rho^{\text{PH}} (s)}{s^n (s - q^2 -\mathrm{i}0^+)} + \sum^{n - 1}_{i = 0} b_i
		(q^2)^i ,
	\end{eqnarray}
where $s_N=36m_Q^2$ is the normal threshold and $b_i(q^2)^i$ is a subtraction term. The hadronic spectral function can be parametrized by using the ``narrow resonance'' assumption
	\begin{eqnarray}
    \begin{split}
        \rho^{\text{PH}} (s) &= \sum_X \delta (s - m_X^2) \langle 0|J|X \rangle \langle X | J^\dagger | 0 \rangle 
        \\&=f_X^2 \delta (s - m_X^2) + \cdots,
    \end{split}
	\end{eqnarray}
in which $m_X$ and $f_X$ are the hadron mass and coupling of the ground state respectively, and \textquotedblleft$\cdots$\textquotedblright \, represents contributions from the continuum and excited states.  

To evaluate the correlation function at the QCD side, we use the $d$-dimensional heavy quark propagator in the momentum space
	\begin{eqnarray}\label{propagator}
    \begin{split}
		  S_Q^{i j} (p) &= \frac{\mathrm{i} \delta^{i j}}{\slashed{p} - m_Q} 
          + \frac{\mathrm{i} \delta^{i j}}{3d} \langle
		g_s^2 G^2 \rangle m_Q \frac{p^2 + m_Q \slashed{p}}{(p^2 - m_Q^2)^4} 
        \\
          &-
		\frac{\mathrm{i}}{4} g_s \frac{\lambda^a_{ij}}{2} G^a_{\mu \nu} \frac{\sigma^{\mu
				\nu} \left( \slashed{p} + m_Q \right) + \left( \slashed{p} + m_Q \right)
			\sigma^{\mu \nu}}{(p^2 - m_Q^2)^2} ,
        \end{split}
	\end{eqnarray}
where $\lambda^a_{ij} \, (a=1,\cdots,8)$ is the Gell-Mann matrix, and $i, j$ are  color indices.

In general, the contribution of the dimension-six triple-gluon condensate is negligible in fully heavy multiquark systems~\cite{Wu:2021tzo,Chen:2016jxd,Wang:2021mma,Wang:2021taf,Yang:2021zrc,Wang:2020avt}. In this work, we calculate the correlation function via the operator product expansion (OPE) method up to dimension-four gluon condensate 
\begin{eqnarray}
    \Pi^{\text{OPE}}(q^2) = \Pi^{\text{pert}}(q^2) + \langle g_s^2 G^2 \rangle \Pi^{\langle g^2_s G^2 \rangle}(q^2).
\end{eqnarray}
\begin{figure}[htbp]
    \centering
    \subfigure[]{
        \includegraphics[width=0.14\textwidth]{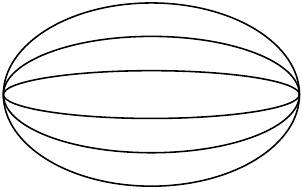}
        \label{fig:Feynman_pert}
    }
    \subfigure[]{
        \includegraphics[width=0.14\textwidth]{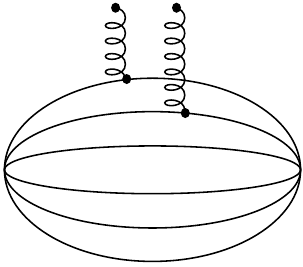}
        \label{fig:Feynman_GG1}
    }
    \subfigure[]{
        \includegraphics[width=0.14\textwidth]{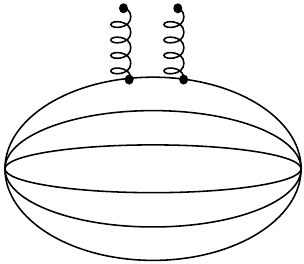}
        \label{fig:Feynman_GG2}
    }
    \caption{Feynman diagrams involved in the OPE series. }
    \label{fig:Feynman}
\end{figure}

As shown in Fig.~\ref{fig:Feynman_pert}-\ref{fig:Feynman_GG2}, the OPE calculations for fully heavy hexaquark correlation function involve massive five-loop banana diagrams, which is an extremely tedious and resource-consuming task. A standard strategy is to reduce these integrals to a linear combination of master integrals via Integration-by-Parts (IBP) identities, which are then computed individually. However, for massive five-loop banana diagrams, the IBP reduction itself becomes prohibitively difficult. Furthermore, it yields an enormous set of master integrals to present the formidable computation challenge. Consequently, the more prevalent technique in QCD sum rules calculations is a sequential, loop-by-loop application of the Feynman parameterization method. While this approach can successfully access the discontinuity (spectral density), it is plagued by very slow numerical integration.

A more efficient alternative is the IDR method. Its central idea is to compute the dispersion representation for a one-loop bubble integral. In this method, the factor $(s-q^2)^{-1}$ effectively acts as a propagator for the next loop integration. This mechanism enables a recursive, loop-by-loop construction of multi-loop integrals. We refer to Refs.~\cite{Remiddi:2016gno,Chen:2024ppj} for a detailed discussion.

While the IDR method easily computes the perturbative term in Fig.~\ref{fig:Feynman_pert}, it encounters difficulties with the two-gluon condensate contributions in Fig.~\ref{fig:Feynman_GG1}-\ref{fig:Feynman_GG2}. The issue arises because these terms involve high-power propagators in Eq.~\eqref{propagator}, which cause a small-circle divergence if the standard dispersion relation is applied to the one-loop bubble computation~\cite{Chen:2024ppj}.

We define the function $B(q^2)$ as an integral to illustrate the structure of small-circle divergence in the massive scalar bubble 
\begin{eqnarray}
    B(q^2)\equiv \int \frac{\mathrm{d}^dk}{\mathrm{i}\pi^{\frac{d}{2}}} \frac{1}{D_0^{n_0}D_1^{n_1}},
\end{eqnarray}
where $D_0=(q-k)^2-m_0^2$, $D_1=k^2-m_1^2$, and $n_0, n_1$ are the powers of the corresponding propagators. Computing the imaginary part, $\operatorname{Im}B(s)$, is typically straightforward. Applying a dispersion relation to this imaginary part then yields the desired dispersion representation for the function $B(q^2)$:
\begin{eqnarray}
\nonumber  B(q^2)=\int_{s_N}^{\infty} \frac{\mathrm{d}s}{\pi} \frac{\operatorname{Im}B(s)}{s-q^2}\left( \frac{q^2}{s} \right)^{n_\Gamma}  + \sum_{k=0}^{n_\Gamma-1} \frac{B^{(k)}(0)}{k!}(q^2)^k , \\
\end{eqnarray}
in which $n_\Gamma$ is the number of subtractions needed, $s_N=(m_0+m_1)^2$ is the physical threshold, and $B^{(k)}(0)$ denotes the $k$-th derivative of $B(q^2)$ at the subtraction point $q^2=0$. Given the asymptotic behavior $\operatorname{Im}B(s) \propto (s-s_N)^\kappa$ near the threshold $s_N$, the dispersion integral converges if $\kappa \geq -\frac{1}{2}$, making the application of the DR valid. When $\kappa < -\frac{1}{2}$, the integral is clearly divergent. 

In QCD sum rules calculations, the problem of small-circle divergence has often not been correctly addressed~\cite{Wang:2022jvk,Wang:2021xao}, as the regularization method adopted yields incorrect results. To properly handle this divergence, one approach involves explicitly computing the small-circle contribution within the framework of standard DR, as demonstrated in Refs.~\cite{Ioffe:2002be,Palameta:2017ols,Esau:2019hqw,Palameta:2018yce}. An alternative is to abandon standard DR in favor of the GDR method introduced in Ref.~\cite{Chen:2024ppj}:
\begin{eqnarray}
    \begin{split}
         B(q^2)&=\int_{s_N}^{\infty} \frac{\mathrm{d}s}{\pi} \frac{\operatorname{Im}B(s)}{s-q^2}\left( \frac{q^2}{s} \right)^{n_\Gamma}  \left( \frac{s-s_N}{q^2-s_N} \right)^{n_\gamma} +
         \\
         & \sum_{k=0}^{n_\Gamma-1} \frac{(q^2)^k}{k!(q^2 -s_N)^{n_\gamma}} \left. \left[(q^2-s_N)^{n_\gamma}B(q^2)\right]^{(k)} \right|_{q^2=0}
         ,
    \end{split}
\end{eqnarray}
where $n_\gamma$ is the number of subtractions needed on the small circle around the point $s=s_N$. This conclusion extends to tensor integrals. In multi-loop calculations, one must check for numerical divergence in the dispersion integral. Given the general complexity, we adopted a different strategy that avoids the GDR calculations entirely, as presented in Ref.~\cite{Xu:2025oqn}.

To simplify computations and numerical analyses, we reduce the correlators $\Pi^{\text{pert}}(q^2)$ and $\Pi^{\langle g^2_s G^2 \rangle}(q^2)$ to be dimensionless as the following
\begin{eqnarray}
\begin{split}
\Pi^{\text{pert}}(q^2) &\equiv m_Q^{14} \tilde{\Pi}^{\text{pert}}(\tilde{q}^2)
\\
\Pi^{\langle g^2_s G^2 \rangle}(q^2) &\equiv m_Q^{10} \tilde{\Pi}^{\langle g^2_s G^2 \rangle}(\tilde{q}^2),
\end{split}
\end{eqnarray}
in which $\tilde{q}^2=q^2/m_Q^2$ is also dimensionless. 
The Borel transform of $\Pi^{\text{pert}}(q^2)$ is denoted as 
\begin{eqnarray}
\begin{split}
    \hat{B}\Pi^{\text{pert}}(q^2) &\equiv  \lim_{\substack{ -q^2, n\to \infty, \\-q^2/n = M_B^2}} \frac{(-q^2)^{n+1}}{\Gamma(n+1)} \left(\frac{\mathrm{d}}{\mathrm{d} q^2}\right)^n \Pi^{\text{pert}}(q^2)
    \\
&=    m_Q^{14} \hat{B}\tilde{\Pi}^{\text{pert}}(\tilde{q}^2)
    \\
    &= m_Q^{16} \int_{\tilde{s}_N}^\infty  \tilde{\rho}^{\text{pert}}(\tilde{s}) \mathrm{e}^{-\tilde{s}/\tilde{M}_B^2 } \mathrm{d} \tilde{s}
\end{split}
\end{eqnarray}
where $M_B^2$ is the Borel mass, $\tilde{s}_N= s_N/m_Q^2$ and $\tilde{M}_B^2 = M_B^2/m_Q^2$ are also reduced. The Borel transform of $\Pi^{\langle g^2_s G^2 \rangle}(q^2)$ can be derived analogously.

Based on the quark-hadron duality, one can establish the QCD sum rules by performing the Borel transform on both $\Pi^{\text{OPE}}(q^2)$ and $\Pi^{\text{PH}}(q^2)$
\begin{eqnarray}
\begin{split}
    \mathcal{L}_k(s_0,M_B^2) &\equiv f_X^2(m_X^2)^k \mathrm{e}^{-m_X^2/M_B^2} 
    \\
    &= m_Q^{16+2k} \int_{\tilde{s}_N}^{\tilde{s}_0}   \tilde{s}^k \tilde{\rho}^{\text{OPE}}(\tilde{s}) \mathrm{e}^{-\tilde{s}/\tilde{M}_B^2 } \mathrm{d} \tilde{s},
    \\
    \tilde{\rho}^{\text{OPE}}(\tilde{s}) &= \tilde{\rho}^{\text{pert}}(\tilde{s}) + m_Q^{-4} \langle g_s^2G^2 \rangle \tilde{\rho}^{\langle g^2_s G^2 \rangle}(\tilde{s}),   \label{eq:dimensionless}
\end{split}
\end{eqnarray}
where $\tilde{s}_0 = s_0/m_Q^2$ is the reduced continuum threshold parameter. The mass of the hadronic ground state can be determined as
\begin{eqnarray} \label{hadronmass}
    m_X(s_0,M_B^2) = \sqrt{\frac{\mathcal{L}_1(s_0,M_B^2)}{\mathcal{L}_0(s_0,M_B^2)}},
\end{eqnarray}
which is the function of $s_0$ and $M_B^2$. 

The dimensionless spectral densities $\tilde{\rho}^{\text{pert}}(\tilde{s})$ and $\tilde{\rho}^{\langle g^2_s G^2 \rangle}(\tilde{s})$ can be obtained by assuming $m_Q = 1$ in the OPE series, which are formally identical for the fully-charm ($cccccc$) and fully-bottom ($bbbbbb$) dibaryon systems. The physical distinction between the two systems enters only through the explicit value of $m_Q$ in the original OPE expression. This dimensionless formulation not only simplifies the analysis but also provides a clear demonstration of how the quark mass governs the overall behavior of the OPE series.

\section{Numerical analyses}  \label{sec3}
In this section, we perform QCD sum rules analyses for the $\Omega_{ccc}\Omega_{ccc}$ and $\Omega_{bbb}\Omega_{bbb}$ dibaryon states by using the following values of heavy quark masses and gluon condensate~\cite{ParticleDataGroup:2024cfk,Narison:2018dcr,Narison:2025cys,Wu:2021tzo,Albuquerque:2016znh,Wu:2022qwd}
\begin{equation}
	\begin{aligned}
m_c^{\overline{\text{MS}}}(m_c) &=1.2730\pm 0.0046 \,\mathrm{GeV}, \\  
m_c^{\text{OS}} & =1.46\pm0.07\ \mathrm{GeV}, \\
m_b^{\overline{\text{MS}}}(m_b) & =4.183\pm 0.007 \,\mathrm{GeV}, \label{eq:parameters} \\  
m_b^{\text{OS}} & =4.65\pm0.05\ \mathrm{GeV} \\
		\langle \alpha_s G^2\rangle & =  (6.35 \pm 0.35)\times 10^{-2}\,\mathrm{GeV}^4,
	\end{aligned}
\end{equation}
in which the heavy quark masses are considered under both the $\overline{\text{MS}}$ and on-shell (OS) renormalization schemes. 
\begin{figure}[]
    \centering
        \includegraphics[width=0.45\textwidth]{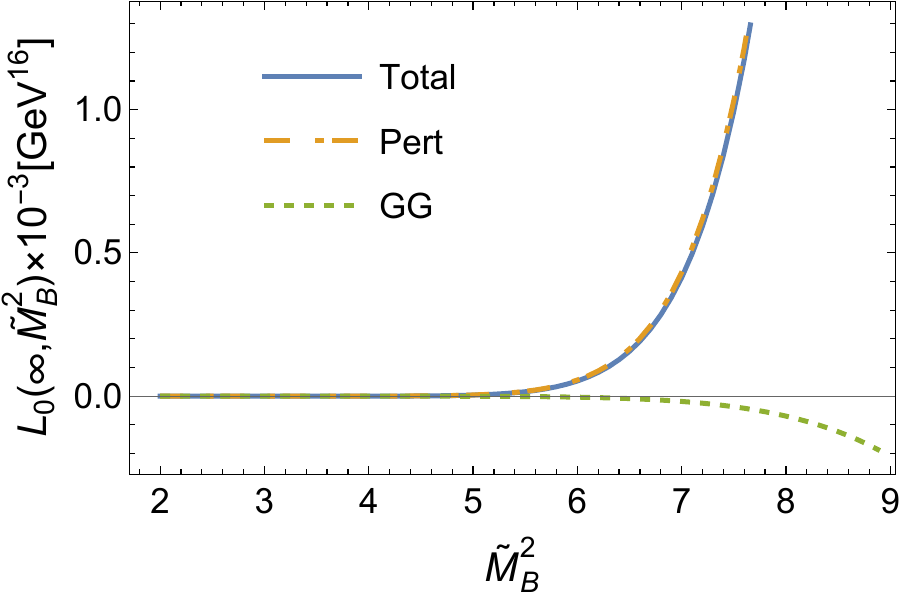}
    \caption{OPE convergence for the $\Omega_{ccc}\Omega_{ccc}$ dibaryon with $J^{P}=0^{+}$.}
    \label{fig:convergence}
\end{figure}

 In the following, we show the analyses for the $\Omega_{ccc}\Omega_{ccc}$ dibaryon with $J^P=0^+$ to demonstrate our analysis. As shown in Eq.~\eqref{hadronmass}, the extracted hadron mass is a function of the continuum threshold parameter $s_0$ and Borel mass $M_B^2$. It is thus important to find suitable parameter space of ($s_0, \, M_B^2$) by studying the behavior of OPE series, pole contribution and the mass sum rules stability.  To ensure the OPE convergence, we require that the contribution of gluon condensate $\langle g^2_s G^2\rangle$ term be less than $30\%$ of the perturbative term
\begin{equation}
    \frac{m^{12}_Q \langle g^2_s G^2 \rangle \int_{36}^\infty \mathrm{e}^{-\tilde{s}/\tilde{M}^2_B} \rho^{\langle g^2_s G^2\rangle}(\tilde{s})\mathrm{d}\tilde{s}}{m^{16}_Q  \int_{36}^\infty \mathrm{e}^{-\tilde{s}/\tilde{M}^2_B} \rho^{\text{pert}}(\tilde{s})\mathrm{d}\tilde{s}} \le 30 \%.  \label{eq:convergence}
\end{equation}
\begin{figure}[]
    \centering
        \includegraphics[width=0.45\textwidth]{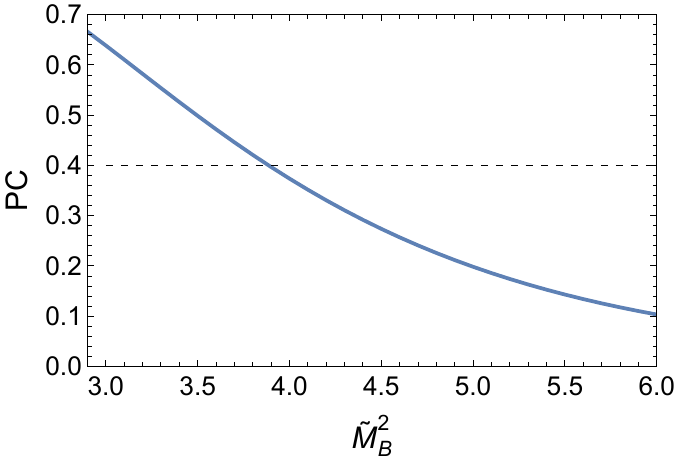}
    \caption{The pole contribution for the scalar $\Omega_{ccc} \Omega_{ccc}$ dibaryon.}
    \label{fig:pole contribution}
\end{figure}
\begin{figure}[]
    \centering
        \includegraphics[width=0.45\textwidth]{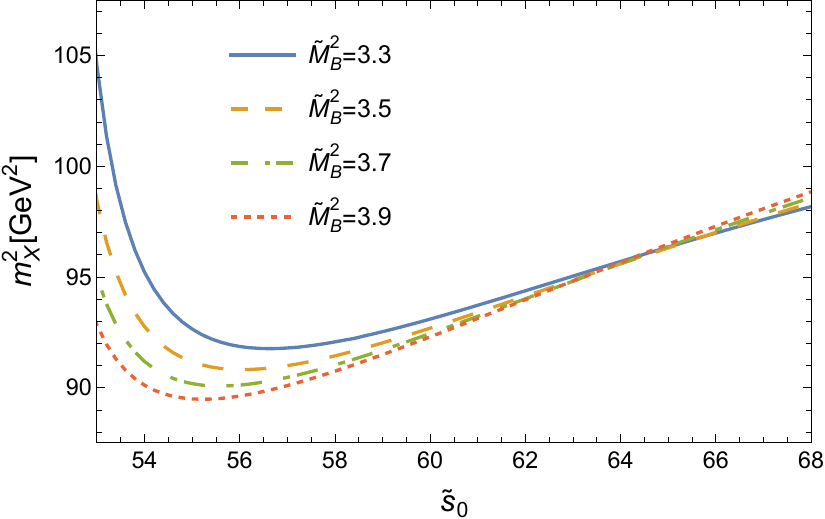}
    \caption{Variation of hadron mass $m_X$ with $\tilde{s}_0$ for the $\Omega_{ccc}\Omega_{ccc}$ dibaryon with $J^{P}=0^{+}$.}
    \label{fig:choice of s0}
\end{figure}
\begin{figure}[]
    \centering
        \includegraphics[width=0.45\textwidth]{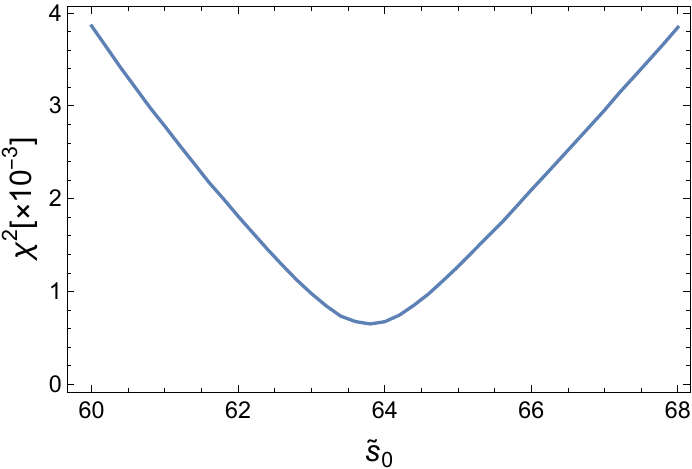}
    \caption{The $\chi^2(\tilde{s}_0)$ for the $\Omega_{ccc}\Omega_{ccc}$ dibaryon with $J^{P}=0^{+}$.}
    \label{fig:chi function}
\end{figure}
As shown in Fig.~\ref{fig:convergence}, the perturbative term gives dominant contribution to the correlation function, providing a lower limit for the Borel mass. 
The upper limit for the Borel mass is established by ensuring that the pole contribution (PC) 
\begin{equation}
    \text{PC}(s_0,M_B^2)=\frac{\mathcal{L}_0(s_0,M_B^2)}{\mathcal{L}_0(\infty,M_B^2)}\ge 40\%. \label{eq:PC}
\end{equation}
We show the variation of pole contribution with the Borel parameter in Fig.~\ref{fig:pole contribution}.
\begin{figure}[]
    \centering
        \includegraphics[width=0.45\textwidth]{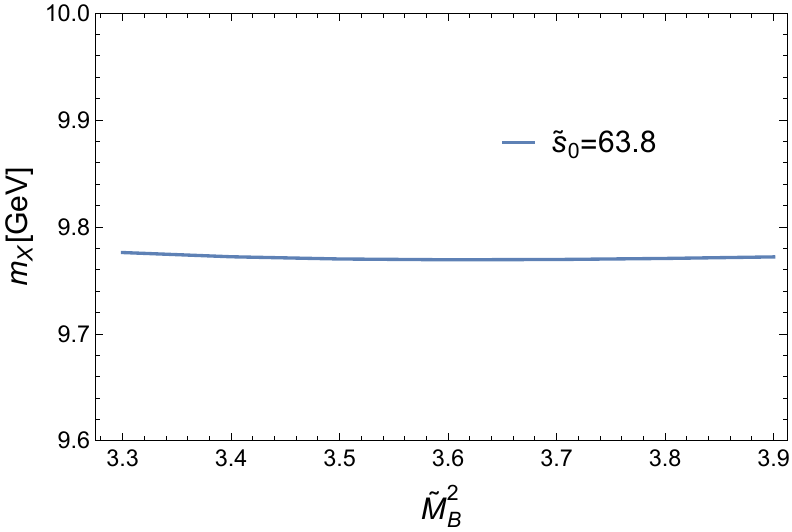}
    \caption{The variation of hadron mass with $\tilde{M}^2_B$ for the scalar $\Omega_{ccc} \Omega_{ccc}$ dibaryon.}
    \label{fig:result of mass}
\end{figure}

To study the mass sum rules stability, we plot the mass curves of $m_X -  \tilde{s}_0$ for different values of $\tilde{M}^2_B$ in Fig.~\ref{fig:choice of s0}. We need to determine the optimal value of the continuum threshold $\tilde{s}_0$, around which the dependence of $m_X$ on $\tilde{M}^2_B$ is minimum. We define $\chi^2$ function as~\cite{Xu:2025oqn,Chen:2024bpz,Li:2025dkw,Li:2025hsp}
\begin{equation}
\chi^2(\tilde{s}_0)=\sum_{i=1}^N\left[\frac{m_X(\tilde{s}_0,\tilde{M}_{B,i}^2)}{\overline m_X(\tilde{s}_0)}-1\right]^2 , \label{eq:chi}
\end{equation}
where $\overline{m}_X(\tilde{s}_0)$ is the average of data points
\begin{equation}
    \overline m_X(\tilde{s}_0)=\sum_{i=1}^N\frac{ m_X(\tilde{s}_0,\tilde{M}_{B,i}^2)}{N},
\end{equation}
in which $\tilde{M}^2_{B,i}(i=1,2,\dots,N)$ are arbitrary points throughout the Borel window. As shown in Fig.~\ref{fig:chi function}, the minimum of $\chi^2(\tilde{s}_0)$ is reached around $\tilde{s}_0=63.8$, for which value the stability of dibaryon mass sum rules can be guaranteed. 

Within the determined parameter space of ($\tilde{s}_0, \, \tilde{M}_B^2$), we show the mass curve for the scalar $\Omega_{ccc}\Omega_{ccc}$ dibaryon in Fig.~\ref{fig:result of mass}, yielding the mass prediction
\begin{equation}
m_X=(9.77 \pm 0.04) \, \text{GeV},
\end{equation}
in which the error comes from the uncertainties of charm quark mass $m_c^{\overline{\text{MS}}}$, the continuum threshold parameter $\tilde{s}_0$ and gluon condensate $\langle g^2_s G^2 \rangle$. It is worth mentioning that the error analysis is greatly facilitated by the dimensionless expression in Eq.~\eqref{eq:dimensionless}. Considering the good stability of mass sum rules in Fig.~\ref{fig:result of mass},  the uncertainty from the Borel mass $\tilde{M}^2_B$ is small enough to be neglected. For the continuum threshold parameter $\tilde{s}_0$, the optimal value is determined by using Eq.~\eqref{eq:chi}. In the numerical analyses, however, we sample $\tilde{s}_0$ on a discrete grid with a step size of $0.2$, yielding a $\pm0.2$ discretization uncertainty of this parameter. 

\begin{figure}[]
    \centering
        \includegraphics[width=0.45\textwidth]{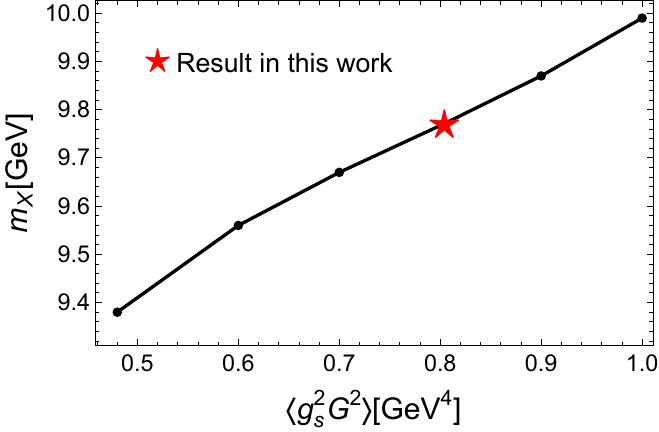}
    \caption{The sensitivity of extracted hadron mass to gloun condensate $\langle g^2_s G^2 \rangle$ for the scalar $\Omega_{ccc}\Omega_{ccc}$ dibaryon.}
    \label{fig:sensitivity of G}
\end{figure}


We also study the tensor $\Omega_{ccc}\Omega_{ccc}$ dibaryon with $J^P=2^+$. Using the $\overline{\text{MS}}$ charm quark mass, the predicted dibaryon mass is around $10.45$ GeV, which is much heavier than that in the scalar channel. Similar QCD sum rules studies can be also extended to the fully bottom $\Omega_{bbb}\Omega_{bbb}$ dibaryon systems with $J^P=0^+$ and $2^+$, by replacing $m_c^{\overline{\text{MS}}}$ to $m_b^{\overline{\text{MS}}}$ in the numerical  analyses. It is found that the sum rules can provide reliable mass predictions with good OPE convergences and pole contributions. The numerical results are collected in Table~\ref{table:MS result} for both $\Omega_{ccc}\Omega_{ccc}$ and $\Omega_{bbb}\Omega_{bbb}$ systems by using the $\overline{\text{MS}}$ heavy quark masses. For the $\Omega_{bbb}\Omega_{bbb}$ systems, the reduced parameters $\tilde{s}_0$ and $\tilde{M}^2_B$ are much smaller than those for the $\Omega_{ccc}\Omega_{ccc}$ systems due to $m_b\gg m_c$.


\begin{table}[htbp] 
	\centering
	\renewcommand\arraystretch{1.4}
	\setlength{\tabcolsep}{1.8mm}
	{
		\begin{tabular}{cccccc}
			\hline\hline
			System & $J^{P}$ &  $\tilde{s}_0(\pm0.2)$ & Mass[$\mathrm{GeV}$]  & $\tilde{M}^2_B$ & $\text{PC}[\%]$\\
			\hline
			$\Omega_{ccc}\Omega_{ccc}$ & $0^{+}$ & 63.8 & $9.77 \pm 0.04$ & 3.3-3.9 & $>40$\\
			$\Omega_{ccc}\Omega_{ccc}$ & $2^{+}$ & 73.6 & $10.45 \pm 0.04$ & 3.9-4.4 & $>40$\\
			$\Omega_{bbb}\Omega_{bbb}$ & $0^{+}$ & 41.8 & $26.60 \pm 0.05$ & 0.57-0.85 & $>40$\\
			$\Omega_{bbb}\Omega_{bbb}$ & $2^{+}$ & 43.2 & $27.05 \pm 0.06$ & 0.64-0.83 & $>40$\\
			\hline\hline
		\end{tabular}
	}
	\caption{Numerical results for the $\Omega_{ccc}\Omega_{ccc}$ and $\Omega_{bbb}\Omega_{bbb}$ dibaryons in $\overline{\text{MS}}$ scheme.}\label{table:MS result}
\end{table}

In the $\overline{\text{MS}}$ scheme, our prediction of the scalar $\Omega_{ccc}\Omega_{ccc}$ dibaryon mass is in  consistent with the quark model calculations of Refs.~\cite{Huang:2020bmb,Weng:2022ohh}, while a bit different from those in Refs.~\cite{Liu:2021pdu,Lyu:2021qsh,Mathur:2022ovu,Wang:2022jvk,Lu:2022myk,Alcaraz-Pelegrina:2022fsi,Martin-Higueras:2024qaw,Dhindsa:2025gae}. For the mass of scalar $\Omega_{bbb}\Omega_{bbb}$ dibaryon, our prediction is much lower than the results in literatures. We show the comparison of these theoretical calculations in Table~\ref{table:comparison with others}.
The fully heavy baryons remain unobserved, yet numerous theoretical calculations exist for the $\Omega_{ccc}$ and $\Omega_{bbb}$ ground-state masses. These predictions, however, show a spread of several hundred MeV. 
The most recent LQCD calculations predicted their masses to be $4793(5)(^{+11}_{-8})$ MeV and $14366(7)(9)$ MeV for the lowest $\Omega_{ccc}$ and $\Omega_{bbb}$ states~\cite{Dhindsa:2024erk,Mathur:2022ovu}, respectively. Accordingly, our calculations in the $\overline{\text{MS}}$ scheme show that the scalar $\Omega_{ccc}\Omega_{ccc}$ dibaryon is slightly higher than the mass threshold of $2\Omega_{ccc}$, while the bound states of $\Omega_{bbb}\Omega_{bbb}$ dibaryons may exist.



It is meaningful to study the dependence of our result on the value of gluon condensate, since it is the only nonperturbative parameter in our calculations.  
To date, there has been various investigations on the gluon condensate $\langle g^2_s G^2\rangle$~\cite{Shifman:1978bx,Reinders:1984sr,Narison:2018nbv,Narison:2025cys,Horsley:2012ra,Chakraborty:2014aca,Geshkenbein:2003cs,Guberina:1980fn,DELPHI:1994tlg}. The early QCD sum rules determined a relative small value $\langle \frac{\alpha_s}{\pi} G^2\rangle \simeq0.012\,\mathrm{GeV}^4$~\cite{Shifman:1978bx}, which is only 60\% of that in Eq.~\eqref{eq:parameters}. In Fig.~\ref{fig:sensitivity of G}, it is clear that the extracted hadron mass is very sensitive to gluon condensate.

It is useful to perform the numerical analyses using on-shell heavy quark masses, given that the spectral functions have been evaluated in the on-shell scheme. We reanalyze the above investigations and collect the numerical results in Table~\ref{table:on-shell result}. Comparing with Table~\ref{table:MS result}, we find that the predicted masses for both the $\Omega_{ccc}\Omega_{ccc}$ and $\Omega_{bbb}\Omega_{bbb}$ states are considerably heavier than those obtained in the $\overline{\text{MS}}$ scheme. Furthermore, the corresponding uncertainties are significantly larger, owing to the sizable errors in the  $\overline{\text{MS}}$ masses $m_{b/c}^{\overline{\text{MS}}}$. Consequently, the mass of the $\Omega_{bbb}\Omega_{bbb}$ state predicted in the on-shell scheme is in good agreement with results from other theoretical methods, whereas the predicted mass of the $\Omega_{ccc}\Omega_{ccc}$ state is somewhat heavier.

\begin{table}[htbp] 
    \centering
    \renewcommand\arraystretch{1.4}
    \setlength{\tabcolsep}{1.8mm}
    {
        \begin{tabular}{c | c c}
            \hline\hline
        Method & $\Omega_{ccc}\Omega_{ccc}\,(\mathrm{MeV})$ &  $\Omega_{bbb}\Omega_{bbb}\,(\mathrm{MeV})$ \\
        \hline
     Lattic QCD ~\cite{Lyu:2021qsh}  & $9585.52
     $ &  / \\
     Lattic QCD ~\cite{Mathur:2022ovu,Dhindsa:2025gae}  & $\Delta E=-39(27)$ &  $28651(^{+16}_{-17})(15)$\\
     Quark Model ~\cite{Martin-Higueras:2024qaw} & $9621.09$ & $28825.62$ \\
     Quark Model  ~\cite{Huang:2020bmb} & $9735.1 \pm 2.7$ & $29780.6 \pm 4.2$ \\
     Quark Model ~\cite{Weng:2022ohh} & $9684.8$ & $28680.7$ \\
     Quark Model ~\cite{Lu:2022myk} & $9960$ & $29167$ \\
     Quark Model ~\cite{Alcaraz-Pelegrina:2022fsi} & $9904$ & $29114$ \\
     OBE ~\cite{Liu:2021pdu} & $  9585.7 $ & $ 28736.3$ \\
     This work($\overline{\text{MS}}$ scheme) &  $ 9770 \pm 40 $  & $ 26600 \pm 50$ \\
This work (OS scheme) &  $ 10840 \pm 460 $ & $ 29230 \pm 310$ \\
            \hline\hline
        \end{tabular}
    }
    \caption{Predicted masses of $\Omega_{ccc}\Omega_{ccc}$ and $\Omega_{bbb}\Omega_{bbb}$ with $J^P=0^+$ in different theoretical approaches. $\Delta E$ means the binding energy. }  \label{table:comparison with others}
\end{table}
 \begin{table}[htbp] 
	\centering
	\renewcommand\arraystretch{1.4}
	\setlength{\tabcolsep}{1.8mm}
	{
		\begin{tabular}{cccccc}
			\hline\hline
			System & $J^{P}$ &  $\tilde{s}_0(\pm0.2)$ & Mass[$\mathrm{GeV}$]  & $\tilde{M}^2_B$ & $\text{PC}[\%]$\\
			\hline
			$\Omega_{ccc}\Omega_{ccc}$ & $0^{+}$ & 59.6 & $ 10.84 \pm 0.46 $ & 2.6-3.3 & $>40$\\
			$\Omega_{ccc}\Omega_{ccc}$ & $2^{+}$ & 65.4 & $ 11.40 \pm 0.47 $ & 3.1-3.4 & $>40$\\
			$\Omega_{bbb}\Omega_{bbb}$ & $0^{+}$ & 40.2 & $ 29.23 \pm 0.31 $ & 0.50-0.59 & $>40$\\
			$\Omega_{bbb}\Omega_{bbb}$ & $2^{+}$ & 41.8 & $ 29.73 \pm 0.32 $ & 0.55-0.65 & $>40$\\
			\hline\hline
		\end{tabular}
	}
	\caption{Numerical results for the masses of $\Omega_{ccc}\Omega_{ccc}$ and $\Omega_{bbb}\Omega_{bbb}$ in on-shell scheme.}\label{table:on-shell result}
\end{table}

\section{Summary}\label{sec4}
In this work, we study the fully heavy dibaryons $\Omega_{ccc}\Omega_{ccc}$ and $\Omega_{bbb}\Omega_{bbb}$ with $J^P=0^+$ and $2^+$ in the method of QCD sum rules. We compose a symmetric tensor dibaryon interpolating current in the molecular configuration, which can couple to both scalar and tensor states.  
We calculate the two-point correlation functions and spectral functions up to dimension four gluon condensate. We use the IDR method to compute the massive five-loop banana diagrams, by properly employing an approach that circumvents the complexity of GDR calculations to deal with the small-circle divergence problem in the gluon condensate term.

The numerical analyses of all fully-heavy dibaryon systems provide stable mass sum rules to give reliable mass predictions. Both in the $\overline{\text{MS}}$ and on-shell scheme, our results show that  the mass of scalar dibaryon is lower than the tensor one for both charm and bottom sectors. In the $\overline{\text{MS}}$ scheme, the scalar $\Omega_{ccc}\Omega_{ccc}$ dibaryon is predicted to be slightly above the $2\Omega_{ccc}$ mass threshold. The masses of $\Omega_{bbb}\Omega_{bbb}$ dibaryons are much lower than $2\Omega_{bbb}$, suggesting the possible existence of bound state in the bottom system. However, the predicted masses in the  on-shell scheme are much heavier for both the $\Omega_{ccc}\Omega_{ccc}$ and $\Omega_{bbb}\Omega_{bbb}$ states.

\textit{\textbf{Acknowledgement.}} \textemdash \,
Xu-Liang Chen and Jin-Peng Zhang thank Xiao-Hui Chen for valuable discussions. This work is supported by the National Natural Science Foundation of China under Grant No. 12575153. 


\end{document}